\documentclass[12pt,a4paper,fleqn]{article}

\usepackage[textheight=23cm,textwidth=15cm]{geometry}
\usepackage{amsmath}
\usepackage{graphicx}
\usepackage{booktabs}
\usepackage{subcaption}
\usepackage{multirow}
\usepackage[american]{babel}
\usepackage{here}
\usepackage[articletitle=true, super=true]{achemso}

\begin{document}

\begin{center}

{\LARGE\bf
A Hybrid Message-Passing and Electric-Field Representation for Electrostatic QM/MM Embedding with Machine Learning Interatomic Potentials
}

\vspace{1cm}

{\large
Raphael T.~Husistein$^{a,}$\footnote{ORCID: 0009-0005-2696-4574},
Shobhit S.~Chaturvedi$^{a,}\footnote{ORCID: 0000-0003-4977-5600}$,
Thomas Weymuth$^{a,}\footnote{ORCID: 0000-0001-7102-7022}$,
Beatriz Piniello$^{b,d,}\footnote{ORCID: 0000-0001-5557-3146}$, 
S\'{e}l\`{e}ne Forget$^{b,d,}\footnote{ORCID: 0009-0002-6006-4041}$
Kresten Lindorff-Larsen$^{b,}$\footnote{ORCID: 0000-0002-4750-6039},
Gemma C.~Solomon$^{c,d,}$\footnote{ORCID: 0000-0002-2018-1529},
and
Markus Reiher$^{a,}$\footnote{Corresponding author; e-mail: mreiher@ethz.ch; ORCID: 0000-0002-9508-1565}
}\\[4ex]

$^{a}$ETH Zurich, Department of Chemistry and Applied Biosciences, Vladimir-Prelog-Weg 2, 8093 Zurich, Switzerland.

$^{b}$University of Copenhagen, Dept.~of Biology, Linderstr{\o}m-Lang Centre for Protein Science, DK-2200, Copenhagen, Denmark.

$^{c}$University of Copenhagen, Dept.~of Chemistry and Nano-Science Center, DK-2100, Copenhagen, Denmark.

$^{d}$University of Copenhagen, Niels Bohr Institute, NNF Quantum Computing Programme, DK-2100, Copenhagen, Denmark.

October 30, 2026

\vspace{.43cm}

\textbf{Abstract}
\end{center}
\vspace*{-.41cm}
{\small
Within a hybrid quantum mechanics/molecular mechanics approach a small region of a large atomistic system is described quantum mechanically, while the rest of the system is represented by a classical force field. Despite this, the computational cost of the quantum calculation severely limits the accessible simulation and sampling time. Replacing the quantum calculation with a machine learning interatomic potential (MLIP) trained on quantum reference data can greatly increase the accessible simulation time. However, the accuracy and efficiency strongly depend on the specific way in which the MLIP incorporates information about the classical (electrostatic) environment. Here, we demonstrate that existing strategies based purely on features derived from an external electric field fail to distinguish certain environments, ultimately limiting their accuracy. As a solution, we propose an extension to the equivariant TensorNet architecture that mitigates this shortcoming by incorporating classical atoms close to the quantum region into the message-passing scheme, combined with features derived from the electric field for describing the environment farther away. We show that this combination achieves higher accuracy than either approach alone.
}

\section{Introduction}
\label{sec:introduction}
While the fundamental physical laws governing essentially all of chemistry and material science are generally viewed as known\cite{Reiher2015}, (approximately) solving these equations turns out to incur costs that scale steeply with the system size (i.e., the number of atoms). There are many approaches that make the study of large systems tractable. They range from molecular mechanics (MM)\cite{Hollingsworth2018}, which treats the system under study classically, through semi-empirical methods\cite{Thiel2014, Husch2018, Bannwarth2021}, which introduce approximations to the quantum mechanical equations, to, more recently, machine learning interatomic potentials (MLIPs)\cite{Jacobs2025, Behler2026, Husistein2026}, which reproduce the quantum mechanical reference data they have been trained on at a reduced computational cost. 

Due to the still comparatively large computational cost of semi-empirical methods and MLIPs for configurational sampling, the default approach for modeling large biomolecular systems comprising tens of thousands of atoms or more remains classical force-field-based molecular mechanics (MM). However, in cases where the process of interest is local (i.e., restricted to a rather small region of a larger system) and/or potentially complex in terms of the interactions involved, such as the active site of a protein, embedding techniques combining quantum-mechanical (QM) methods with classical force fields, usually denoted QM/MM, can be used. The small region is then described quantum mechanically, whereas the remainder of the system is treated classically\cite{Senn2009, Csizi2023}. If high quantum accuracy is key, but sampling cannot be ignored, the advent of MLIPs has opened up the way to replace the QM method by an MLIP leading to machine learning/molecular mechanics (ML/MM) schemes\cite{Grassano2025}.

There exist different options for performing QM/MM calculations. In the simplest case, the so-called mechanical embedding, the QM calculation is performed in the absence of the MM region and the interaction is treated at the MM level. As a consequence, the polarization of the QM region by the MM region is neglected. By contrast, in electrostatic embedding, the MM point charges are included in the QM Hamiltonian and thereby allow the MM region to polarize the QM region. In biomolecular simulations, electrostatic embedding is the most widely used scheme\cite{Senn2009}. 

Various approaches for constructing MLIPs for ML/MM embeddings have been published in the literature and they use different techniques for integrating information about the MM environment into the model. In what is arguably the simplest case, the model never directly sees the environment. Instead, the model predicts the gas-phase energy of the QM region, which is then coupled classically to the MM region. AceFF-2-RESP achieves this by predicting partial charges that are used for calculating the short-range Coulomb interaction with the MM charges (the long-range part is calculated with force-field charges) and by adding an analytical correction for polarization based on fixed atomic polarizabilities\cite{Farr2026}. EMLE\cite{Zinovjev2023}, by contrast, uses a charge equilibration scheme\cite{Rappe1991} to obtain charges used in the calculation of the electrostatic interactions, while polarization is treated with the Thole model\cite{Thole1981}. A different approach is taken by models that incorporate MM atoms within a cutoff directly into the descriptor or the graph\cite{Boeselt2021, Hofstetter2022, Bensberg2025}. Outside the cutoff, however, the MM environment is described either by means of Coulomb interactions with fixed charges, which means that the polarization of the QM region by far-away MM atoms is not captured\cite{Bensberg2025}, or by a semi-empirical method, where the ML model only learns the difference between DFT and the semi-empirical method (called $\Delta$-learning)\cite{Boeselt2021, Hofstetter2022}. Other works learn the QM/MM coupling by supplying information about the electrostatics of the MM environment as input to the MLIP. NepoIP\cite{Song2025} exploits the electrostatic potential at each QM atom as a scalar feature, while FieldSchNet\cite{Gastegger2021} is coupling the electric field vector to learned dipole features. However, both approaches are blind to MM environments with vanishing electrostatic potential or field at the QM atoms, respectively (see also Section \ref{subsec:symmetry-probes} below). FieldMACE\cite{Barrett2026} partially addresses this limitation by extending the MACE architecture to expand the potential generated by the MM charges in spherical harmonics, up to an arbitrary order, and then using the expansion coefficients as features during message passing. However, any approach relying on a truncated expansion remains blind to environments where the expansion vanishes up to truncation order. A fourth approach, implemented in the anisotropic message passing (AMP) model\cite{Pultar2025}, uses predicted multipoles of the QM atoms that interact explicitly with the MM point charges, while polarization is described using learned induced multipoles. A drawback of this method is that it requires not only energies and forces but also molecular multipoles for training. Furthermore, MM-induced multipoles only enter in the last message-passing step which limits the extent to which MM atoms can alter the features of QM atoms.

In order to address these shortcomings, we developed a new approach, combining two different strategies. Similar to FieldMACE, we exploit local electric field features (electrostatic potential, electric field vector, electric field gradient) at each QM atom, but we also integrate nearby MM atoms into the message-passing procedure. This allows us to capture the polarization caused by MM atoms outside the cutoff through the field features and, unlike other expansion-based approaches, to distinguish environments in which all local electric field features vanish, since the positions of the MM atoms within the cutoff are resolved explicitly.

This paper is organized as follows: in Section \ref{sec:methods} we detail the TensorNet architecture as well as our extensions specifically designed for treating QM/MM systems. Next, in Section \ref{sec:results}, we demonstrate that describing the MM environment using only local electric field features leads to degeneracies. We then evaluate the performance of our new approach, which circumvents this issue, using several model systems, before giving a conclusion in Section \ref{sec:conclusion}.

\section{Methods}
\label{sec:methods}
As our electrostatic MLIP is based on the previously published TensorNet\cite{Simeon2023} architecture, we will briefly review its most important architectural components before describing our extensions to make it useful in the context of QM/MM simulations.

\subsection{Base TensorNet}
TensorNet belongs to the class of equivariant graph neural networks\cite{Scarselli2009}. At a high level, it consists of three main modules: embedding, interaction, and output. We will briefly explain the key features of each module, but we omit details such as the feature normalization. For a more detailed description we refer the interested reader to Ref.~\citenum{Simeon2023}.

In the first step, the molecule or atomistic assembly will be converted into a graph, where edges connect atoms that are less than the cutoff distance $R_c$ apart. The distance between atoms is then converted into a vector by expanding it using radial basis functions, given by $\varphi_i(r_{ij}) = \exp(-\beta_i(\exp(-r_{ij})-\mu_i))^2$ where $\beta_i$ and $\mu_i$ are fixed parameters. To encode the element for each atom, an embedding layer is used, whereby a random vector is initialized for each element and later optimized during training. The geometry is encoded in irreducible Cartesian tensors up to rank 2, which can be stored efficiently in a $3\times3$ matrix. More concretely, for each edge $e_{ij}$ the $3\times3$ matrix is constructed by combining the $l=0$, $l=1$, $l=2$ features. To do so, the scalar (i.e., $l=0$) feature is encoded as multiple of the identity, the vectorial (i.e., $l=1$) feature as skew-symmetric matrix, and the $l=2$ as symmetric traceless tensor. All features are derived from the normalized displacement vector $r_{ij}$ and combined with information about the elements joined by the given edge and the output of a multi-layer perceptron (MLP). The MLP takes the distance encoded via the radial basis functions as input and predicts multiple output values, resulting in several $3\times3$ feature matrices per edge. Finally, node features are calculated by summing all features of the node's edges.

After the embedding module has generated the initial node features, TensorNet usually has several interaction modules stacked sequentially. They refine the initial node features and integrate information from nodes that are more than one hop away. For this purpose, the $3\times3$ features are decomposed into $l=0, 1$, and $2$ components (called the irreducible components) and aggregated in a linear combination with learnable weights (called $Y$). The updated node state ($Y$) is again decomposed into its irreducible components. An MLP, which takes the distance expanded in radial basis functions as input, generates a coefficient for each component. The node message is then computed by the weighted sum of these components. Each node sums the messages of its neighbors (resulting in $M$) before calculating the interaction product given by $I = Y\cdot M + M\cdot Y$. This product is coupling features of different angular momentum similar to the Clebsch--Gordan tensor product used in models based on spherical harmonics. Before the final node update, the interaction product is decomposed into irreducible components once more and a learned linear combination is constructed ($Y'$). The final node feature is 
then given by the sum of the previous node feature and the update $Y' + Y^{\prime 2}$.

The output module leverages the node's features to produce the final prediction. We will focus on scalar predictions for simplicity, but TensorNet can be used to predict tensorial properties as well. In the first step, the node features are decomposed into irreducible components. To make the prediction invariant under rotations, invariant features are extracted for each component by calculating the squared Frobenius norm given by $\operatorname{Tr}(\mathbf{X}^T\mathbf{X})$. Those features are concatenated into a vector and passed through an MLP to get a node-level scalar prediction. In the case of a global prediction such as the total energy, one can simply sum up the node-level predictions to get the final value.

\subsection{Extensions for a QM/MM embedding}
\label{subsec:extensions-for-qmmm}
For QM/MM calculations with electrostatic embedding, the Hamiltonian of the QM region is augmented with an additional term describing the Coulomb interaction with the point charges used to represent the MM atoms\cite{Senn2009}. Therefore, a simple ansatz to train an MLIP with electrostatic embedding would be to pass the full system (i.e., the coordinates and chemical species) of all QM and MM atoms to the MLIP. However, this has the disadvantage that the same amount of compute is spent for QM and MM atoms despite the MM atoms being simple point charges and in turn limits the size of system one can treat. An approach that circumvents that unfavorable scaling is the use of local field features generated by the MM atoms. 

In our case, we focus on the electrostatic potential and its first two derivatives:
\begin{align}
  V(\mathbf{r}) &= \sum_{i}^M \frac{q_i}{|\mathbf{r} - \mathbf{r}_i|} \\
  \mathbf{E}(\mathbf{r}) &= \sum_{i}^M \frac{q_i\,(\mathbf{r} - \mathbf{r}_i)}{|\mathbf{r} - \mathbf{r}_i|^3} \\
  \nabla\mathbf{E}(\mathbf{r}) &= \sum_{i}^M q_i\frac{|\mathbf{r}-\mathbf{r}_i|^2\mathbf{I} - 3(\mathbf{r}-\mathbf{r}_i)(\mathbf{r}-\mathbf{r}_i)^\top}{|\mathbf{r}-\mathbf{r}_i|^5},
\end{align}
where $\mathbf{r}$ is the position of the QM atom, $M$ is the number of MM atoms, and $q_i$ and $\mathbf{r}_i$ are the charge and position of MM atom $i$. We note that these can be classified according to their order as tensors of rank zero (scalar, $l=0$), rank one (vector, $l=1$), and rank two ($l=2$), respectively. To leverage these features, we introduce a new field embedding module positioned inside the embedding module.

To ease capturing non-linear effects, the electrostatic potential $V$ is projected into a higher dimensional space using a Gaussian basis expansion. Concretely, the expansion is given by
\begin{equation}
    g_k(V) = \exp\left(-\frac{(V-\mu_k)^2}{2\sigma^2}\right), \qquad k = 1,\dots,N
\end{equation}
where the centers $\mu_k$ are linearly spaced between the minimal and maximal value of $V$ (specified initially) and sigma is equal to the range of $V$ divided by the number of Gaussians $N$ (a hyperparameter). The result is passed through a linear layer and then integrated into the model with feature-wise linear modulation\cite{Perez2018}. That means the feature vector is split in two halves, $\gamma$ and $\beta$, and then used to modulate the embedding of the chemical species $z$:
\begin{equation}
    \text{FiLM}(z) = \big(1+\tanh(\gamma)\big)\odot z + \tanh(\beta),
\end{equation}
where $\odot$ represents element-wise multiplication and $\tanh$ bounds the modulation terms to $[-1, 1]$. Since these embeddings are then used to construct edge features, FiLM provides a (direction independent) modification of pairwise interactions.

The key ingredient to leverage the electric field vector and its gradient is to encode them in the same way that TensorNet encodes its features: as a skew-symmetric matrix and as a symmetric traceless tensor, respectively. An MLP is used to let the chemical element and the electric field influence the extent to which a node feature is altered by the field. For this purpose, the MLP takes as input the embedding of the chemical element, the Euclidean norm of the field vector, and the Frobenius norm of the field gradient, with both norms bounded to $[0, 1]$ via the $\tanh$ function. For each node feature, it predicts two gate values: one for the field and one for its gradient. These gate values are then multiplied with the field features and the result is added to the node features. 

As demonstrated in Section \ref{subsec:arginine-glutamate}, including these features substantially improves prediction accuracy compared to a model relying solely on QM atoms. However, a model depending only on local electric field features is unable to distinguish certain configurations (see Section \ref{subsec:symmetry-probes}), which motivates the inclusion of MM atoms in message passing. Since QM and MM atoms are described by different physical models, the model must be able to distinguish between the two. This is achieved by using separate element embeddings, allowing the model to learn different roles for each atom type. Since the model predicts only the energy of the QM region, only the features of the QM atoms are passed on to the output module. The MM atoms can therefore influence the energy by altering the features of the QM atoms, but they do not directly contribute to the energy. Because including all MM atoms in the message-passing procedure would be prohibitively expensive, only MM atoms that are within  $R_{\mathrm{MM}}$ of a QM atom are included in the graph. In contrast, the local electric field features are calculated from all MM atoms, allowing the model to account for long-range electrostatic effects of the whole environment. To make sure that the energy does not display any discontinuities when an MM atom is no longer included in the graph, the contribution of MM atoms is smoothly damped, starting at $R_{\mathrm{MM}} - \Delta_{\mathrm{MM}}$ and reaching zero at $R_{\mathrm{MM}}$. This is achieved by using a shifted version of the cosine cutoff function\cite{Behler2011}: 
\begin{equation}
c(d) =
\begin{cases}
1, & d \le R_{\mathrm{MM}} - \Delta_{\mathrm{MM}}, \\[4pt]
\dfrac{1}{2}\left[1 + \cos\!\left(\pi\cdot\,\dfrac{d - R_{\mathrm{MM}} + \Delta_{\mathrm{MM}}}{\Delta_{\mathrm{MM}}}\right)\right], & R_{\mathrm{MM}} -
\Delta_{\mathrm{MM}} < d < R_{\mathrm{MM}}, \\[10pt]
0, & d \ge R_{\mathrm{MM}},
\end{cases}
\end{equation}
where $d$ is the distance of an MM atom to its nearest QM atom. The distance cutoff of each edge is then multiplied by $c(d)$ of both atoms connected by the edge, with $c(d)=1$ for QM atoms. We note that the virtual sites of the optimal point charge (OPC) water model\cite{Izadi2014}, which are massless point charges without an associated element, are included in the local electric field features but not in the graph.

This combines the advantage of the accurate geometrical description enabled by the equivariant message passing, with the benefits of a cheaper description of the full MM system by means of local electric field features.

\subsection{Long-range Electrostatics}
\label{subsec:long-range-electrostatics}
Since our model is based on a fixed cutoff radius, capturing long-range interactions needs information to propagate through many hops. While this primarily involves long-range QM--MM electrostatic interactions, in the case of multiple QM regions that are separated by a distance larger than the cutoff, as for example in the umbrella sampling of Arginine--Glutamate (see Section \ref{subsec:arginine-glutamate} below), it may also involve the interaction between two QM fragments. Additionally, while the overall error for these configurations might be small, a distance-dependent error might lead to a systematic bias. To account for that, we added corrections for the long-range electrostatics. Concretely, we use force-field charges to subtract from the reference energies and forces the Coulomb interaction between the QM and MM regions, and, in cases where the QM region contains two separate fragments $\text{QM}_1$ and $\text{QM}_2$, the interaction between those fragments. The MLIP is trained on the remainder and both terms are added back during inference. The energy terms are given by:
\begin{align}
  E_\text{QM-MM} &= \sum_{i \in \text{QM}}\sum_{j \in \text{MM}} \frac{q_i q_j}{\left|\mathbf{r}_i - \mathbf{r}_j\right|} \\
  E_\text{QM-QM} &= \sum_{i \in \text{QM}_1}\sum_{j \in \text{QM}_2} \frac{q_i q_j\, s\!\left(\left|\mathbf{r}_i - \mathbf{r}_j\right|\right)}{\left|\mathbf{r}_i - \mathbf{r}_j\right|} \\
  s(r) &=
    \begin{cases}
        0, & r < 1.5\,\text{Å} \\
        6t^5 - 15t^4 + 10t^3, & 1.5\,\text{Å} \le r < 2.0\,\text{Å} \\
        1, & r \ge 2.0\,\text{Å}
    \end{cases}
    \qquad t = \frac{r - 1.5\,\text{Å}}{0.5\,\text{Å}}
\end{align}
where $q_i$ are the force-field charges and $s(r)$ is the smootherstep function\cite{Perlin2002} with vanishing first and second derivatives at both ends.

\section{Results and Discussion}
\label{sec:results}
As described in Section \ref{subsec:extensions-for-qmmm}, our extension of TensorNet allows for the inclusion of the external electrostatic potential and its first two derivatives, and additionally MM atoms within $R_{MM}$ of a QM atom. Before tackling a biomolecular test system, namely a capped arginine--glutamate pair in explicit water, we will motivate our choice of features by studying highly symmetric charge distributions.

\subsection{Symmetry Probes}
\label{subsec:symmetry-probes}
To illustrate the benefits of including higher-order field features and incorporating (some) MM atoms in the message passing procedure, we construct charge environments where specific field features vanish. All energies shown in the figures below were calculated with the CCSD(T)\cite{Raghavachari1989} implementation in ORCA 6.1\cite{Neese2025} and the basis set aug-cc-pVQZ\cite{Woon1993}.

\begin{figure}
    \centering
    \begin{subfigure}[t]{\linewidth}
        \centering
        \includegraphics[width=\linewidth]{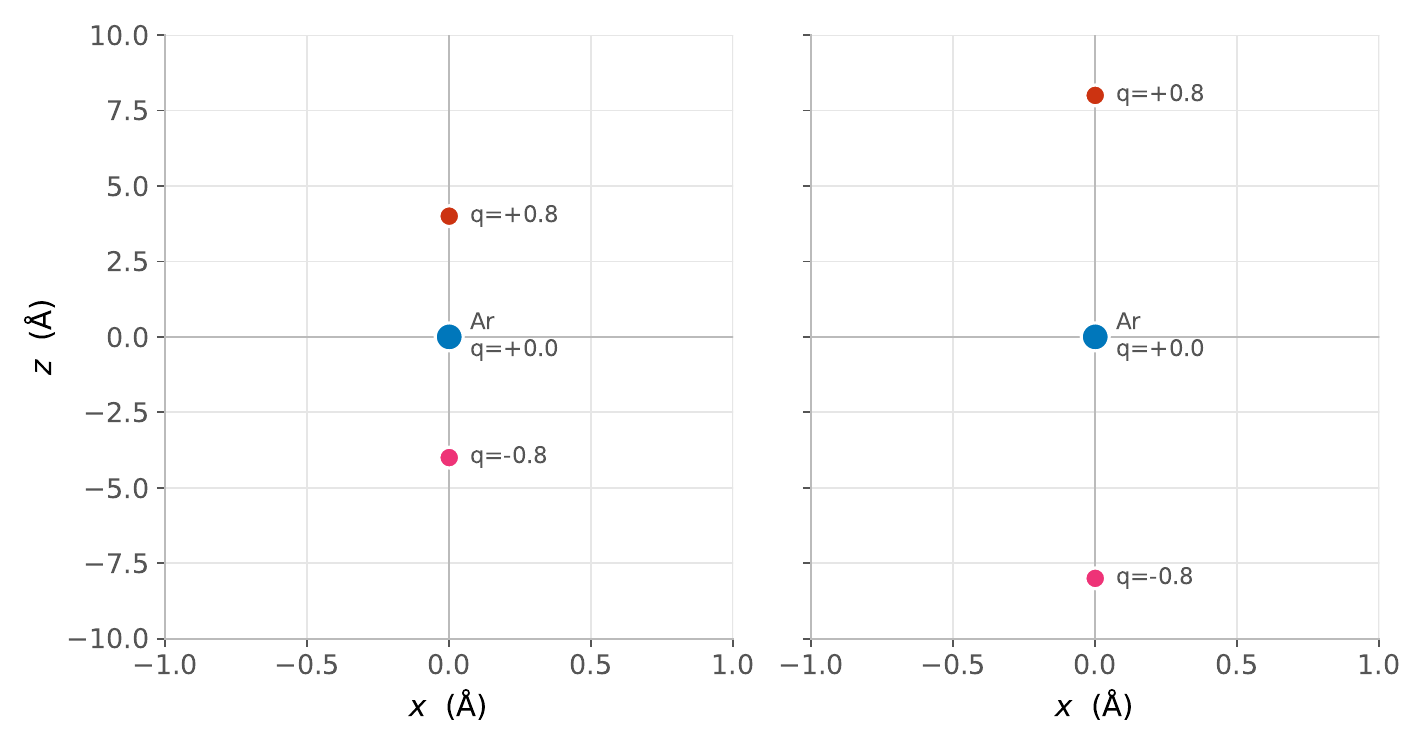}
        \caption{Example geometries for two values of $d$.}
        \label{fig:system-a}
    \end{subfigure}
    \hfill
    \begin{subfigure}[t]{\linewidth}
        \centering
        \includegraphics[width=\linewidth]{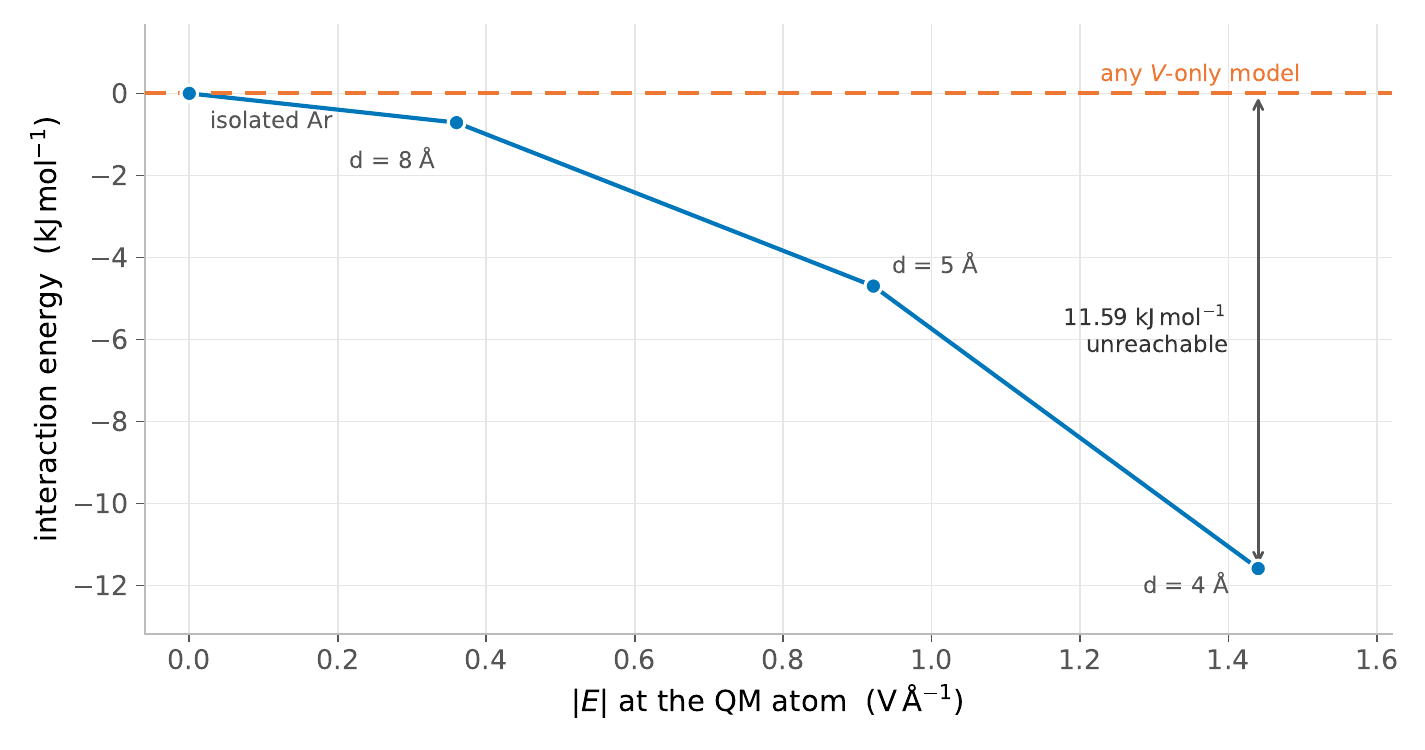}
        \caption{CCSD(T) interaction energy against $|E|$.}
        \label{fig:energy-a}
    \end{subfigure}
    \caption{An argon atom at the origin with opposite charges placed symmetrically at $\pm d$. This symmetric placement cancels $V$ at the origin, and in turn any model solely relying on $V$ will be unable to reproduce the energy curve.}
    \label{fig:probe-a}
\end{figure}

As the simplest case, we consider just a single QM atom and two charges of opposite sign placed at $\pm d$ (Fig.~\ref{fig:system-a}). While the interaction energy strongly depends on the distance between the QM atom and the charges (Fig.~\ref{fig:energy-a}), the electrostatic potential is zero at the position of the QM atom, independent of the distance. This highlights the fact that an MLIP relying on the electrostatic potential as a feature alone is unable to learn the correct energies even for simple cases.

\begin{figure}
    \centering
    \begin{subfigure}[t]{\linewidth}
        \centering
        \includegraphics[width=\linewidth]{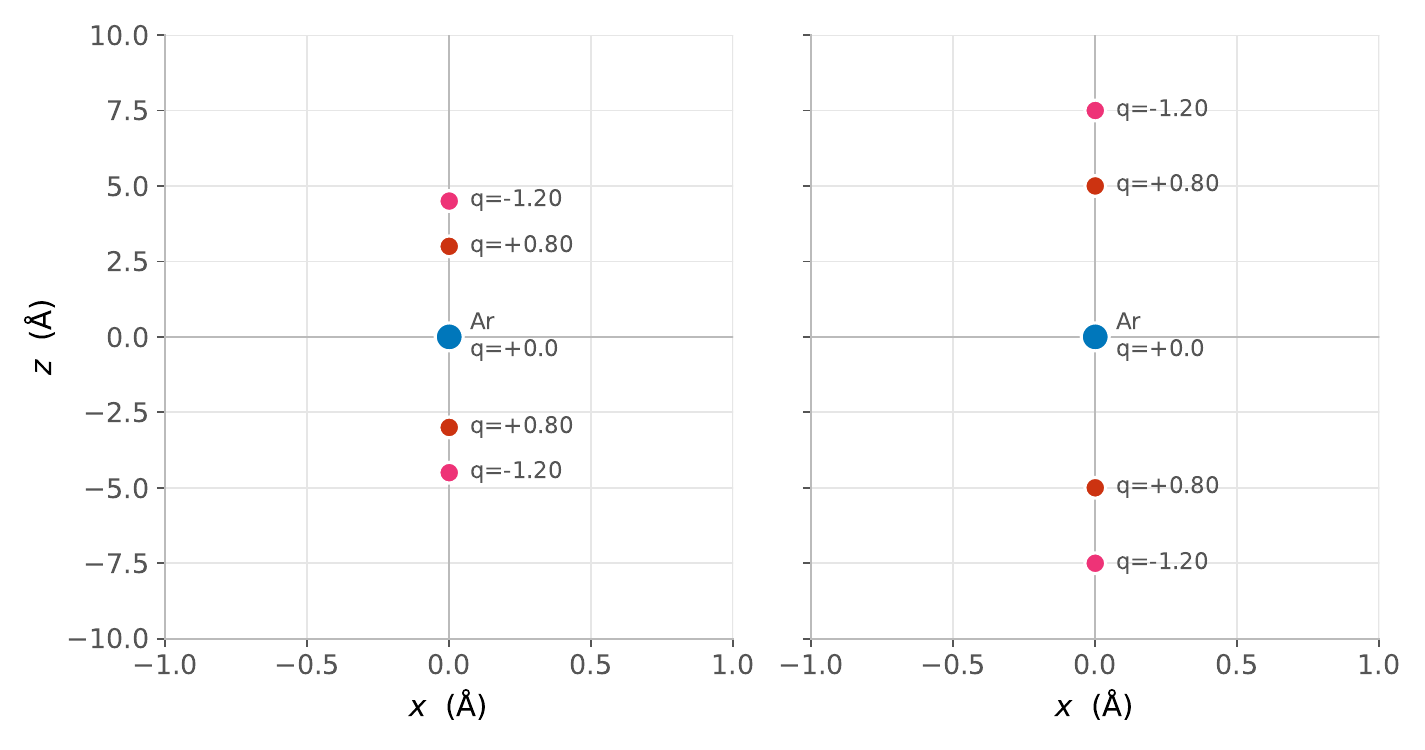}
        \caption{Example geometries for two values of $d_1$ and $d_2$.}
        \label{fig:system-b}
    \end{subfigure}
    \hfill
    \begin{subfigure}[t]{\linewidth}
        \centering
        \includegraphics[width=\linewidth]{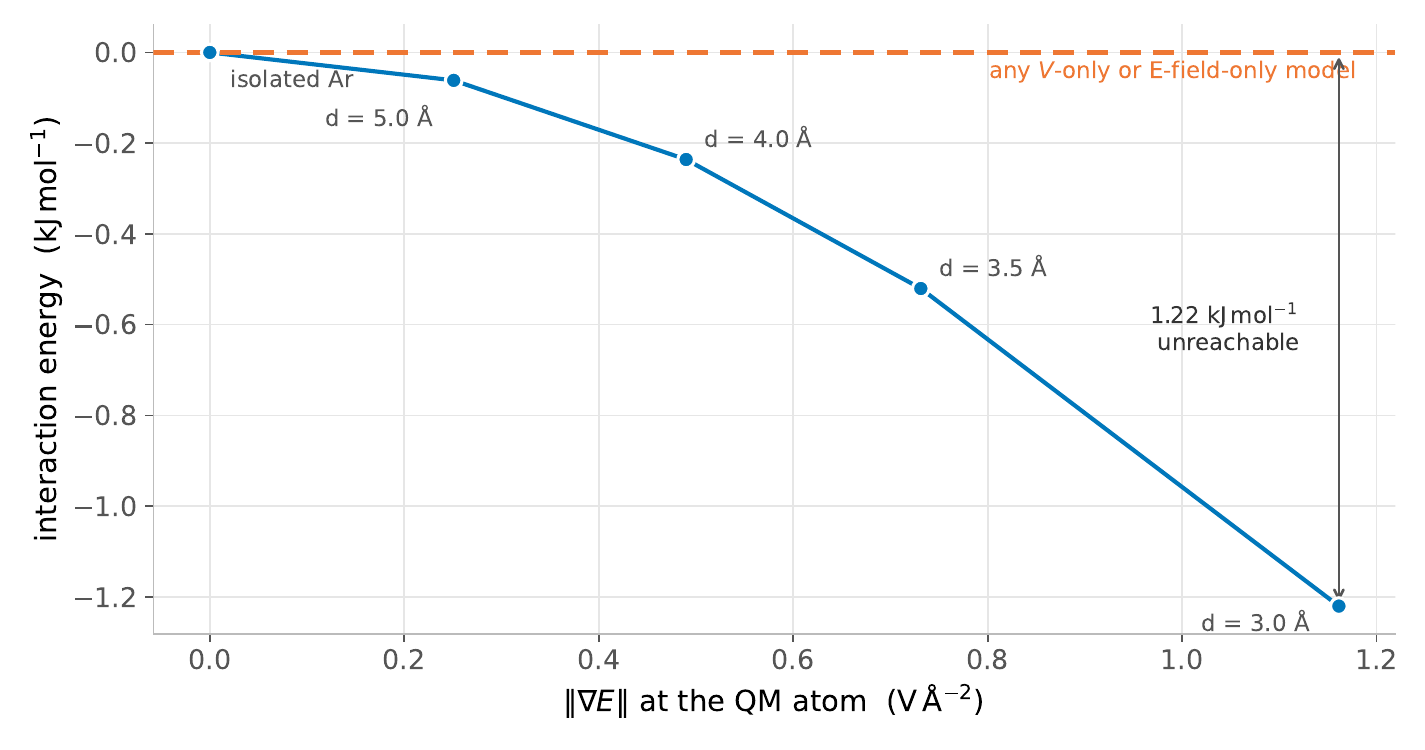}
        \caption{CCSD(T) interaction energy against $||\nabla E||$.}
        \label{fig:energy-b}
    \end{subfigure}
    \caption{An argon atom at the origin with a pair of positive charges at $\pm d_1$ and a pair of negative charges at $\pm d_2$. This symmetric placement cancels $V$ and $\lVert \mathbf{E} \rVert$ at the origin, and in turn any model solely relying on those quantities will be unable to reproduce the energy curve.}
    \label{fig:probe-b}
\end{figure}

By placing two pairs of charges, one pair positively charged and one pair negatively charged, at $\pm d_1$ and $\pm d_2$ from the QM atom, respectively (Fig.~\ref{fig:system-b}), we can construct a system where the electrostatic potential and the electric field vector are both zero at the origin. Since all four charges lie on the $z$-axis, $E_x=E_y=0$. To see that $E_z=0$ as well, notice that the charge distribution has a mirror symmetry through the plane $z=0$ and thus $V(0, 0, z) = V(0, 0, -z)$. By taking the derivative with respect to $z$ on both sides $V'(0, 0, z) = -V'(0, 0, -z)$ and plugging in $z=0$ one arrives at $2V'(0, 0, 0) = 0$ and hence $E_z = -V' = 0$. Although the energy depends on $d_1, d_2$ (Fig.~\ref{fig:energy-b}), a model considering only the electrostatic potential and the electric field vector is unable to distinguish these charge arrangements. By contrast, a model that takes the electric field gradient into account is able to distinguish these configurations.

\begin{figure}
    \centering
    \begin{subfigure}[t]{\linewidth}
        \centering
        \includegraphics[width=\linewidth]{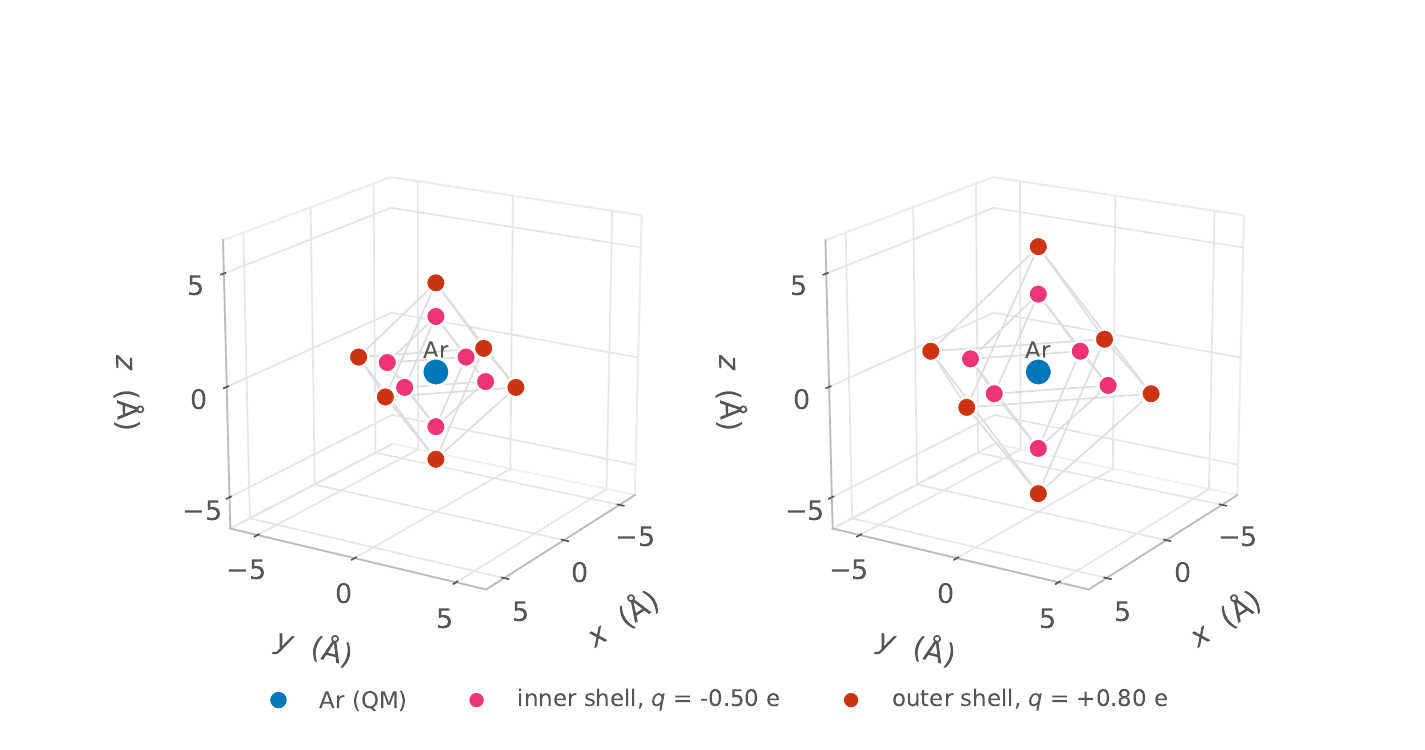}
        \caption{Example geometries for two scalings of the nested octahedra.}
        \label{fig:system-c}
    \end{subfigure}
    \hfill
    \begin{subfigure}[t]{\linewidth}
        \centering
        \includegraphics[width=\linewidth]{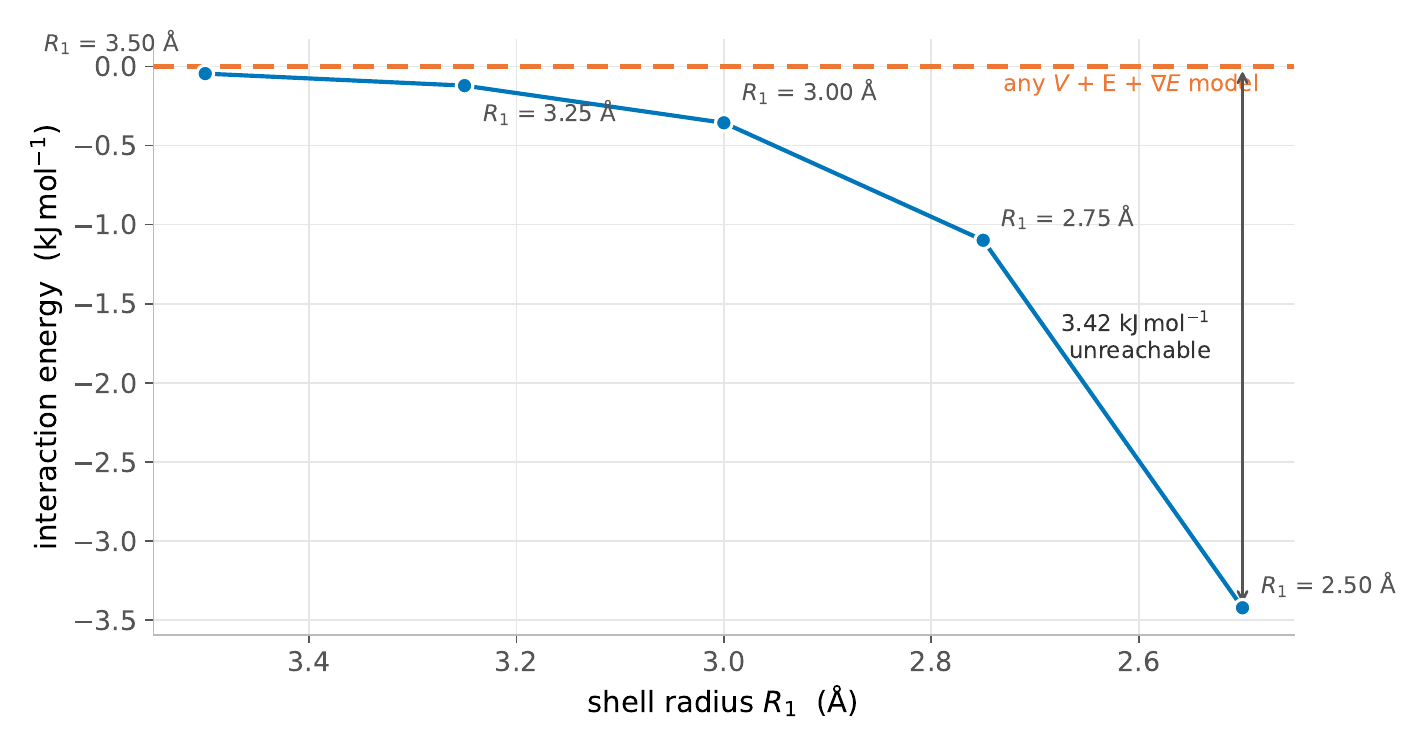}
        \caption{CCSD(T) interaction energy against the distance from the origin to the inner octahedron.}
        \label{fig:energy-c}
    \end{subfigure}
    \caption{An argon atom at the origin surrounded by two octahedra of charges with different sign and magnitude. The symmetric placement cancels $V$, $\lVert \mathbf{E} \rVert$, and $\lVert \nabla\mathbf{E} \rVert$ at the origin and in turn any model solely relying on those quantities will be unable to reproduce the energy curve.}
    \label{fig:probe-c}
\end{figure}

Of course, there are also charge arrangements that make the electric field gradient vanish. For example, placing six charges around the argon atom at the origin in an octahedral configuration makes the electric field gradient at the origin zero. By nesting two such octahedra carrying charges of opposite sign (Fig.~\ref{fig:system-c}), the external potential and all its derivatives up to third order ($l=3$) are zero at the origin, while the energy depends on the distance between the argon atom and the charges (Fig.~\ref{fig:energy-c}). While one could resolve this by integrating higher-order features, there are always arrangements to make these zero. For instance, nesting two spherically symmetric charge distributions (with suitably chosen charges) leads to a vanishing electrostatic potential and all its derivatives. By contrast, including MM atoms within a certain cutoff in the message-passing scheme allows the MLIP to distinguish these highly symmetric configurations (when the charges lie within the cutoff) while still leveraging the electrostatic information from all MM atoms, including atoms beyond the cutoff.

\subsection{Arginine--Glutamate}
\label{subsec:arginine-glutamate}
In the following, we show that these features can be used to systematically improve the accuracy of an MLIP trained on a real QM/MM system, namely a capped arginine--glutamate pair in explicit water. The training data consisted of $8,000$ structures, which were extracted from trajectories obtained through enhanced molecular dynamics simulation using the umbrella sampling technique. A more detailed description of the procedure can be found the Supporting Information.% \ref{si:all_atom}. 
The side-chain center-of-mass (scCOM) distance between the two pairs varies from $1.8\mathrm{nm}$ to $0.3\mathrm{nm}$, divided into $18$ windows. Additionally, $931$ structures sampled from all 18 windows were used as test set. The QM/MM reference data were calculated using the r$^2$SCAN\cite{Furness2020, Furness2020b} exchange--correlation functional augmented with D3 dispersion correction\cite{Grimme2010}, Becke–-Johnson damping (BJ)\cite{Grimme2011} and def2-TZVP basis set\cite{Weigend2005}. To correct for basis set superposition errors, we used the Boys and Bernardi counterpoise correction\cite{Boys1970}.

The $8,000$ structures were randomly split into training ($90\,\%$) and validation ($10\,\%$) sets. The models were optimized using the AdamW\cite{Loshchilov2019} optimizer with an initial learning rate of $3.4\times10^{-4}$. We used a learning rate schedule, halving the learning rate if validation loss plateaued for $15$ epochs and training stopped once progress on the validation loss stalled for $50$ consecutive epochs. All the hyperparameters used can be found in the Supporting Information. % \ref{si:model-architecture}.

Evaluating the various models on the test set confirms the expectations set by the analysis performed in Section \ref{subsec:symmetry-probes} (see Table \ref{tab:ablation-results}). The model using only QM atoms as input exhibits the highest energy and force error. Performance systematically improves with the inclusion of additional higher-order field features. While incorporating MM atoms into message passing improves performance, the best results are achieved by combining this approach with local field features. We note that the best model shows a mean absolute error well below $4.18\,\mathrm{kJ \cdot mol^{-1}}$, which is often considered to be the threshold for chemical accuracy.

\begin{table}
\centering
\caption{Effect of electronic and MM structural features on test set performance. The results are averaged over three random seeds. The force error is given in units of $\mathrm{kJ \cdot mol^{-1} \cdot \text{\AA}^{-1}}$ and the energy in $\mathrm{kJ \cdot mol^{-1}}$. $\mathbf{N}_\text{atoms}$ is the mean number of atoms used in the model's input graph.}
\label{tab:ablation-results}
\begin{tabular}{llll}
\toprule
\textbf{Configuration} & $\mathbf{N}_\text{atoms}$ & \textbf{Force MAE} & \textbf{Energy MAE} \\
\midrule
QM only & $63$ & $6.44 \pm 0.04$ & $12.12 \pm 0.73$ \\
QM + $V$ & $63$ & $3.76 \pm 0.02$ & $11.05 \pm 0.19$ \\
QM + $V$ + $\mathbf{E}$ & $63$ & $2.45 \pm 0.03$ & $8.06 \pm 0.08$ \\
QM + $V$ + $\mathbf{E}$ + $\nabla^2 V$ & $63$ & $1.92 \pm 0.00$ & $4.69 \pm 0.04$ \\
QM + MM shell & $574$ & $2.25 \pm 0.03$ & $4.32 \pm 0.13$ \\
QM + MM shell + $V$ + $\mathbf{E}$ + $\nabla^2 V$ & $574$ & $1.15 \pm 0.03$ & $1.94 \pm 0.12$ \\
\bottomrule
\end{tabular}
\end{table}

As mentioned in Section \ref{subsec:long-range-electrostatics}, we subtract the classical Coulomb electrostatic interaction between the MM region and the two amino acids, as well as between and the two amino acids themselves. Although we achieved chemical accuracy without subtracting the interaction between the two amino acids, we observed a slight bias in the force as the distance between them increased. Given that the MLIP is intended to be used for generating a potential of mean force, we wanted to avoid this bias. For this analysis, we defined the bias as:
\begin{equation}
F_{\text{bias},k} = \frac{1}{N_k} \sum_{n \in k} \left( F_{\xi,n}^\text{MLIP} - F_{\xi,n}^\text{DFT} \right),
\end{equation}
where the frames are grouped into bins based on the distance $\xi$ between the centers of mass of the side chains, $N_k$ is the number of frames in bin $k$, and $F_{\xi, n}$ is defined as:
\begin{equation}
    F_{\xi,n} = \frac{M_\text{Glu} \mathbf{F}_{\text{Arg},n} - M_\text{Arg} \mathbf{F}_{\text{Glu},n}}{M_\text{Arg} + M_\text{Glu}} \cdot \hat{\mathbf{u}}_n,
\qquad \hat{\mathbf{u}}_n = \frac{\mathbf{R}_{\text{Glu},n} - \mathbf{R}_{\text{Arg},n}}{\left|\mathbf{R}_{\text{Glu},n} - \mathbf{R}_{\text{Arg},n}\right|},
\end{equation}
with $M_\text{Glu}$ being the total mass of the glutamate side chain, $\mathbf{R}_{\text{Glu},n}$ its center of mass, and $\mathbf{F}_{\text{Glu}, n}$ the net force acting on it. The quantities for arginine are defined analogously.

Our analysis (see Table \ref{tab:bias_analysis}) shows that the force bias between the two amino acids can be reduced significantly by subtracting the classical Coulomb interaction between the two amino acids at scCOM--scCOM distances above $15\text{\AA}$. For the model with the explicit Coulomb correction, all confidence intervals contain zero, meaning that we detected no statistically significant bias.

\begin{table}
  \centering
  \caption{Bias of the projected force $F_{\text{bias},k}$ (in kJ\,mol$^{-1}$\,\AA$^{-1}$) per bin for a model trained without and with the explicit Coulomb correction between the two amino acids. The $95\,\%$ confidence intervals (CI) were obtained by bootstrapping. $\xi$ is the bin's centre of the scCOM--scCOM distance and $N_k$ is the number of frames. Highlighted in bold face are the entries whose CI does not contain zero.}
  \label{tab:bias_analysis}
  \begin{tabular}{rrrcrc}
    \toprule
    & & \multicolumn{2}{c}{Without LR correction} & \multicolumn{2}{c}{With LR correction} \\
    \cmidrule(lr){3-4}\cmidrule(lr){5-6}
    $\xi$ (\AA) & $n$ & Bias & 95\% CI & Bias & 95\% CI \\
    \midrule
    3.73 & 79 & $\mathbf{1.30}$ & [$\mathbf{0.51}$, $\mathbf{2.16}$] & $0.55$ & [$-1.12$, $2.00$] \\
    4.87 & 56 & $0.62$ & [$-0.18$, $1.45$] & $-1.13$ & [$-4.22$, $0.55$] \\
    6.02 & 61 & $-0.20$ & [$-1.14$, $0.58$] & $-1.13$ & [$-3.02$, $0.09$] \\
    7.16 & 83 & $\mathbf{0.63}$ & [$\mathbf{0.15}$, $\mathbf{1.09}$] & $0.00$ & [$-0.48$, $0.46$] \\
    8.30 & 81 & $-0.29$ & [$-0.76$, $0.18$] & $0.01$ & [$-0.42$, $0.47$] \\
    9.45 & 81 & $-0.05$ & [$-0.41$, $0.36$] & $-0.04$ & [$-0.44$, $0.35$] \\
    10.59 & 81 & $0.10$ & [$-0.24$, $0.49$] & $-0.05$ & [$-0.41$, $0.30$] \\
    11.73 & 74 & $-0.02$ & [$-0.52$, $0.42$] & $-0.25$ & [$-0.69$, $0.16$] \\
    12.87 & 83 & $0.12$ & [$-0.22$, $0.45$] & $-0.01$ & [$-0.35$, $0.34$] \\
    14.02 & 83 & $-0.06$ & [$-0.52$, $0.44$] & $0.04$ & [$-0.36$, $0.44$] \\
    15.16 & 81 & $\mathbf{-0.80}$ & [$\mathbf{-1.31}$, $\mathbf{-0.36}$] & $-0.01$ & [$-0.33$, $0.34$] \\
    16.30 & 88 & $\mathbf{-1.58}$ & [$\mathbf{-1.99}$, $\mathbf{-1.16}$] & $-0.15$ & [$-0.45$, $0.16$] \\
    \bottomrule
  \end{tabular}
\end{table}

\subsection{Uracil and Benzene in Water}
To demonstrate that our model offers favorable accuracy compared to previous approaches, we compare it on parts of the data set published in Ref.~\citenum{Boeselt2021}, namely benzene and uracil solvated in water. Including the nearby water up to a distance of $5\,\text{\AA}$ from the QM region, the graph for benzene contains on average about 115 atoms, while for uracil it contains about 106 atoms. We follow their protocol and similarly use the first $70\,\%$ of the frames for training while the subsequent $20\,\%$ and $10\,\%$ are used for validation and testing, respectively. For the sake of a fair comparison, we did not subtract the long-range QM-MM electrostatics. Compared to previous work, our model shows improved accuracy for both energy and forces on both data sets (see Table \ref{tab:boeselt_comparison} and \ref{tab:hofstetter_comparison}). Because Ref.~\citenum{Hofstetter2022} reports test set errors only for its $\Delta$-learning models (which learn the difference between DFT and the semi-empirical density functional tight binding (DFTB) method\cite{Elstner1998, Elstner2006}), we compare against its non-$\Delta$-learning models on the validation set. We refrain from a direct comparison with FieldMACE\cite{Barrett2026} because their exact protocol for splitting the data into train and test sets is unknown, and we observe severe data leakage (i.e., near identical frames from the MD landing in the train and test sets) when using random splitting on the uracil data set (see Table \ref{tab:uracil_split_comparison}), demonstrating that a fair comparison is only possible based on matching data splits.

\begin{table}
    \centering
    \caption{Test set performance for uracil and benzene in water given in $\mathrm{kJ \cdot mol^{-1} \cdot \text{\AA}^{-1}}$ and $\mathrm{kJ \cdot mol^{-1}}$. The values for M10 and M4, the best performing non $\Delta$-learning models in Ref.~\citenum{Boeselt2021}, are taken directly from the reference. For our results, we show the mean and the sample standard deviation calculated over three random seeds.}
    \label{tab:boeselt_comparison}
    \begin{tabular}{llcc}
    \toprule
    \textbf{System} & \textbf{Model} & \textbf{Force MAE} & \textbf{Energy MAE} \\
    & & (kJ\,mol$^{-1}$\,\AA$^{-1}$) & (kJ\,mol$^{-1}$) \\
    \midrule
    \multirow{2}{*}{Benzene} & M4\cite{Boeselt2021} & $4.85$ & $4.0$ \\
    & \textbf{Ours} & $\mathbf{1.46 \pm 0.12}$ & $\mathbf{0.55 \pm 0.04}$ \\
    \midrule
    \multirow{2}{*}{Uracil} & M10\cite{Boeselt2021} & $12.4$ & $4.8$ \\
    & \textbf{Ours} & $\mathbf{8.40 \pm 0.52}$ & $\mathbf{3.54 \pm 0.50}$ \\
    \bottomrule
    \end{tabular}
\end{table}

\begin{table}
    \centering
    \caption{Validation set performance for uracil and benzene in water given in $\mathrm{kJ \cdot mol^{-1} \cdot \text{\AA}^{-1}}$ and $\mathrm{kJ \cdot mol^{-1}}$. The values for the GCNN models are taken directly from Ref.~\citenum{Hofstetter2022}. For each system, we compare against the lowest force and energy errors amnong all reported non-$\Delta$-learning GCNN variants. For our results, we show the mean and the sample standard deviation calculated over three random seeds.}
    \label{tab:hofstetter_comparison}
    \begin{tabular}{llcc}
    \toprule
    \textbf{System} & \textbf{Model} & \textbf{Force MAE} & \textbf{Energy MAE} \\
    & & (kJ\,mol$^{-1}$\,\AA$^{-1}$) & (kJ\,mol$^{-1}$) \\
    \midrule
    \multirow{2}{*}{Benzene} & GCNN\cite{Hofstetter2022} & $4.87$ & $2.4$ \\
    & \textbf{Ours} & $\mathbf{0.75 \pm 0.01}$ & $\mathbf{0.31 \pm 0.02}$ \\
    \midrule
    \multirow{2}{*}{Uracil} & GCNN\cite{Hofstetter2022} & $12.97$ & $7.3$ \\
    & \textbf{Ours} & $\mathbf{8.38 \pm 0.43}$ & $\mathbf{3.03 \pm 0.31}$ \\
    \bottomrule
    \end{tabular}
\end{table}

\begin{table}
    \centering
    \caption{Comparison of the error on the uracil test set when using chronological splitting, as in Ref.~\citenum{Boeselt2021}, and when using random splitting.}
    \label{tab:uracil_split_comparison}
    \begin{tabular}{lcc}
    \toprule
    \textbf{Split} & \textbf{Force MAE} & \textbf{Energy MAE} \\
    \midrule
    Chronological & $8.40 \pm 0.52$ & $3.54 \pm 0.50$ \\
    Random        & $0.24 \pm 0.01$ & $0.024 \pm 0.001$ \\
    \bottomrule
    \end{tabular}
\end{table}

\section{Conclusions}
\label{sec:conclusion}
In this work, we extended the equivariant graph neural network TensorNet to make it suitable for training on QM/MM calculations based on electrostatic embedding. Our extension allows for the integration of MM atoms into the message-passing scheme, while the local field features enter the node features. We have demonstrated that both local electric field features and the inclusion of nearby MM atoms in message passing are necessary for the model to be able to distinguish highly symmetric charge distributions and that this translates into measurable accuracy gains when training on QM/MM data for a capped arginine--glutamate pair in water. We have shown that the reliance on a short-range model leads to a force bias between the two amino acids as the distance between the two grows (despite accurate mean absolute forces), and that this can be easily corrected by accounting for the classical Coulomb interaction. This allows the model to be used, for instance, in the context of umbrella sampling for generating a potential of mean force, where this bias would otherwise lead to systematic deviations in the free energy curve. Since only a small proportion of the MM atoms enter the message-passing scheme, and the rest of the environment is treated with local electric field features, it is possible to describe systems larger than what is possible with a traditional MLIP. More broadly, this development is a step toward the use of MLIPs as accurate surrogate models in QM/MM calculations of large biomolecular systems such as enzyme-catalyzed reactions.

\section{Acknowledgments}
This work was created as part of NCCR Catalysis (grant number 225147), a National Centre of Competence in Research funded by the Swiss National Science Foundation.
The authors also gratefully acknowledge financial support from XPRIZE through a Phase-2 finalist award. S.F., B.P. and G.C.S. are supported by the Novo Nordisk Foundation, Grant number NNF22SA0081175, NNF Quantum Computing Programme.

\section*{Author Contributions}
R.T.H.: Design and implementation of QM/MM extension to TensorNet, training of the MLIP, implementation of the automation pipeline (support), and writing (lead).
S.S.C.: Implementation of the automation pipeline (lead), bias analysis, long-range QM-QM correction, QM/MM training data generation, and testing.
T.W.: Implementation of the automation pipeline (support), writing (support).
B.P.: Preparation of arginine and glutamate structures.
S.F.: Preparation of arginine and glutamate structures.
K.L-L., G.C.S., and M.R. planned and supervised the project and acquired the financial funding and compute. 
All authors reviewed and approved the final paper draft.

% references in the SI
\nocite{Tribello2014, Elfwing2018, Tian2020, Sengupta2021, Sengupta2021b, Case2023, Eastman2024}
%\bibliography{references}
\providecommand{\latin}[1]{#1}
\makeatletter
\providecommand{\doi}
  {\begingroup\let\do\@makeother\dospecials
  \catcode`\{=1 \catcode`\}=2 \doi@aux}
\providecommand{\doi@aux}[1]{\endgroup\texttt{#1}}
\makeatother
\providecommand*\mcitethebibliography{\thebibliography}
\csname @ifundefined\endcsname{endmcitethebibliography}
  {\let\endmcitethebibliography\endthebibliography}{}

\section*{Supporting Information}

\subsection*{All-atom Simulation of Arginine--Glutamate}
\label{si:all_atom}
The two system were modeled as a pair of capped amino acids (ACE-X-NME) and built with AmberTools~\cite{Case2023} and OpenMM 8.4~\cite{Eastman2024}. The pair was solvated in a rhombic dodecahedron box with $2.2\,\mathrm{nm}$ padding, with $0.15\,\mathrm{M}$ NaCl (Li/Merz ions \cite{Sengupta2021, Sengupta2021b}) and OPC water\cite{Izadi2014}, yielding a total of $7,623$ atoms. The amino acid elements were modeled using the Amber ff19SB force field \cite{Tian2020}. The all-atom simulations were performed using OpenMM 8.4 \cite{Eastman2024} patched with PLUMED 2.10~\cite{Tribello2014}, with a $2\,\mathrm{fs}$ timestep and with hydrogen-heavy atom bond constraints.

After a $10,000$ step minimization, $2\,\mathrm{ns}$ NVT and $1\,\mathrm{ns}$ NPT restrained equilibration to reach the simulation temperature ($298\,\mathrm{K}$) and pressure ($1\,\mathrm{bar}$), and finally $10\,\mathrm{ns}$ unrestrained NPT production, umbrella sampling was performed along the inter-residue side-chain center-of-mass distance. Eighteen windows were distributed between $0.3$ and $1.8\,\mathrm{nm}$ following a geometric scheme, with spring constants ranging from $1000$ to $6000\,\mathrm{kJ\,mol^{-1}\,nm^{-2}}$ tuned to ensure near-constant overlap between neighboring windows. Sequential seeding started at $1.5\,\mathrm{nm}$ from a $10\,\mathrm{ns}$ unbiased checkpoint and was later extended to $1.8\,\mathrm{nm}$. Each window was simulated for $50\,\mathrm{ns}$, with $3$ independent replicas, yielding a total of $150\,\mathrm{ns}$ of all-atom explicit solvent umbrella sampling of arginine--glutamate.

\subsection*{Model Architecture and Hyperparameters}
\label{si:model-architecture}
Details of the model architecture and training hyperparameters are given in Table~\ref{tab:model-architecture}. The loss was a weighted sum of the mean squared errors of the energies and forces.
\begin{table}
  \centering
  \caption{Model architecture and training hyperparameters.}
  \label{tab:model-architecture}
  \begin{tabular}{lcc}
      \hline
      Parameter & Arg--Glu & Uracil/Benzene \\
      \hline
      Radial cutoff & $6\,\text{\AA}$ & $5\,\text{\AA}$ \\
      MM cutoff $R_{\mathrm{MM}}$ & $6\,\text{\AA}$ & $5\,\text{\AA}$ \\
      MM damping width $\Delta_{\mathrm{MM}}$ & $1\,\text{\AA}$ & $1\,\text{\AA}$ \\
      Field embedding ($V$, $\mathbf{E}$, $\nabla\mathbf{E}$) & yes & yes \\
      Gaussian expansion range for $V$ & $[-10, 10]$ & $[-3, 2]$ \\
      Number of Gaussians for $V$ & $32$ & $16$ \\
      Atomic reference energies & yes & no \\
      Maximum neighbors & $512$ & $64$ \\
      Embedding dimension & $128$ & $128$ \\
      Radial basis functions & $64$ & $64$ \\
      Interaction layers & $3$ & $2$ \\
      Output block dimensions & $512, 256, 128, 64$ & $128, 64$ \\
      Activation function & SiLU\cite{Elfwing2018} & SiLU\cite{Elfwing2018} \\
      Energy loss weight & $1.0$ & $2.0$ \\
      Force loss weight & $100.0$ & $1.0$ \\
      Learning rate & $3.439\times10^{-4}$ & $10^{-4}$ \\
      Weight decay & $10^{-5}$ & -- \\
      Gradient clipping (norm) & $10$ & $40$ \\
      Effective batch size & $32$ & $32$ \\
      Maximum epochs & $3000$ & $3000$ \\
      Early-stopping patience & $50$ epochs & $100$ epochs \\
      LR scheduler patience & $15$ & $20$ \\
      LR scheduler factor & $0.5$ & $0.8$ \\
      \hline
  \end{tabular}
\end{table}

\end{document}